# Surveying College Introductory Physics Students' Attitudes and Approaches to Problem Solving

Andrew J. Mason[1] and Chandralekha Singh[2]
[1]Department of Physics and Astronomy, University of Central Arkansas, Conway, Arkansas, USA. Corresponding author.
[2]Department of Physics and Astronomy, University of Pittsburgh, Pittsburgh, Pennsylvania, USA

**Abstract**
Students' attitudes and approaches to problem solving in physics can greatly impact their actual problem solving practices and also influence their motivation to learn and ultimately the development of expertise. We developed and validated an Attitudes and Approaches to Problem Solving (AAPS) survey and administered it to students in the introductory physics courses in a typical large research university in the US. Here, we discuss the development and validation of the survey and analysis of the student responses to the survey questions in introductory physics courses. The introductory physics students' responses to the survey questions were also compared with those of physics faculty members and physics Ph.D. students. We find that introductory students are in general less expert-like than the physics faculty members and Ph.D. students. Moreover, on some AAPS survey questions, the responses of students and faculty have unexpected trends. Those trends were interpreted via individual interviews, which helped clarify reasons for those survey responses.

## Introduction

Thinking like a scientist amounts to building not only a good knowledge structure of science concepts and developing problem solving and meta-cognitive skills, but also positive attitudes about the knowledge and learning in science [1-2]. Prior studies have emphasized that students' attitudes towards learning and problem solving in science, as well as their conceptions of what it means to learn science [3-8], can have a significant impact on what they actually learn [9-15]. For example, it is impossible to become a true physics expert without a simultaneous evolution of expert-like approaches and attitudes about the knowledge and learning in physics. If students think that physics consists of a collection of disconnected facts and formulas rather than a coherent hierarchical structure of knowledge, they are unlikely to make an effort to organize their knowledge hierarchically. Similarly, if students feel that only a few smart people can do physics, the instructor is the authority and the students' task in a physics course is to take notes, memorize the content and reproduce it on the exam and then forget it, they are unlikely to make an effort to synthesize and analyze what is taught, ask questions about how concepts fit together or how they can extend their knowledge beyond what is taught. Similarly, if students believe that if they cannot solve a problem within 10 minutes, they should give up, they are unlikely to persevere and make an effort to explore strategies for solving challenging problems.

The Maryland Physics Expectation Survey (MPEX) was developed to explore students' attitudes and expectations related to physics [13]. When the survey was administered before and after instruction in various introductory physics courses, it was found that students' attitudes about physics after instruction deteriorated compared to their expectations before taking an introductory physics course. Very few carefully designed courses and curricula have shown major improvements in students' expectations after an introductory physics course [16-17]. The Colorado Learning Assessment Survey (CLASS) is another survey, which is similar to the MPEX survey and explores students' attitudes about physics [9]. The analysis of CLASS data yields qualitatively similar results to those obtained using the MPEX survey.

Cummings, Lockwood, and Marx [17-18] developed the Attitudes towards Problem Solving Survey (APSS), which is partially based upon MPEX. The original APSS survey has 20 questions and examines students' attitudes towards physics problem solving. The survey was given to students before and after instruction at three types of institutions: a large university, a smaller university and a college. It was found that students' attitudes about problem solving did not improve after instruction (in fact, they deteriorated slightly) at the large university and the attitudes were least expert-like (least favorable) at the large university with a large class.

Since students' attitudes and approaches to learning and problem solving can affect how they approach physics learning and how much time they spend learning, instructors should employ explicit strategies to improve them. According to the field-tested cognitive apprenticeship model, students can learn effective problem solving strategies and build a robust knowledge structure if the instructional design involves three essential components: modeling, coaching & scaffolding, and weaning [19]. In this approach, "modeling" means that the instructor demonstrates and exemplifies the skills that students should learn (e.g., how to solve physics problems systematically). "Coaching & scaffolding" means that students receive appropriate guidance and support as they actively engage in learning the skills necessary for good performance. "Weaning" means reducing the support and feedback gradually to help students develop self-reliance.

Schoenfeld was motivated by the cognitive apprenticeship model to develop a curriculum to improve students' attitude about problem solving in mathematics [20-22] since students' beliefs about problem solving in mathematics is similar in many respects to students' beliefs about problem solving in physics. For example, based upon his knowledge that students in the introductory mathematics courses often start looking for formulas right away while solving problems instead of performing a careful conceptual analysis and planning, Schoenfeld used an explicit strategy to change students' problem solving approach. He asked students to work in small groups and solve complex problems. During the group activities, he would move around and ask students questions such as "What are you doing? Why are you doing it? How does it take you closer to your goals?" Students who were used to immediately looking for formulas without even making sense of the problem realized that they should first perform conceptual analysis and planning before jumping into the implementation of the problem solution by formula fitting. Schoenfeld's strategy helped students with novice-like problem solving approaches adopt an effective problem solving approach within a short period of time and they started to devote time in conceptual analysis and planning of the problem solving before looking for equations.

Another unfavorable attitude about mathematical problem solving that Schoenfeld had observed is that students often felt that if they could not solve a problem within 5-10 minutes, they should give up [19-22]. Schoenfeld realized that one reason students had such an unproductive attitude was that they observed their instructor solving problems during the lectures without faltering like them or spending too much time thinking. To remove this unproductive conception about problem solving, Schoenfeld began each of his geometry classes with the first 10 minutes devoted to taking students' questions about challenging geometry problems (often the most challenging end of the chapter exercises) and thus attempting to solve them without prior preparation. Students discovered that Schoenfeld often struggled with the problems and was unable to solve them in 10 minutes and asked students to continue to think about the problems until one of them had solved it and shared it with others. This approach improved students' attitude and raised their self-confidence in being able to solve mathematics problems if they use effective approaches to problem solving and keep trying. Recently, Adams and Wieman attributed the difficulties in assessing problem solving skills to the existence of more than 40 sub-skills [23]. Instructional design based upon the cognitive apprentice model can be helpful in teaching these sub-skills relevant for a knowledge rich domain such as physics to a variety of students [24].

Here, we first summarize the development and validation of the Attitudes and Approaches to Problem Solving (AAPS) survey. We then discuss the responses of students in college introductory physics courses in the US on the AAPS survey, a modified version of the APSS survey [18] that includes additional questions, particularly related to various approaches to problem solving. We analyze how introductory physics students differ in their attitudes and approaches from physics faculty members and Ph.D. students. We find that, on average, the attitudes and approaches of introductory students differ significantly from faculty and Ph.D. students in several ways. On some of the AAPS survey items, the responses of students and faculty have unexpected trends. In order to interpret those trends, we conducted interviews with a subset of individuals to clarify reasons for their responses.

*Validity and Reliability of the Survey*

Below, we summarize issues related to the validity and reliability of the survey before discussing the findings. Validity refers to the appropriateness of interpreting the survey scores [25-26]. The AAPS survey can be found in the Appendix. In order to develop the survey, we selected 16 questions from the APSS survey [18] and further modified some of them for clarity based upon in-depth interviews with five introductory physics students and discussions with some physics faculty members. These 16 questions constitute the first 14 questions and the last two questions of the APSS survey. We also developed 17 additional questions, many of which focused on approaches to problem solving, and modified them based upon the feedback from introductory students during interviews, and discussions with three physics faculty and some Ph.D. students. The introductory students and faculty members were both important for validation purpose at this stage because the responses of these two groups were the most disparate and provided the most diverse feedback for improving the preliminary versions of the survey. Some of the themes in the additional questions are related to the use of diagrams and scratch work in problem solving, use of "gut" feeling vs. using physics principles to answer conceptual questions, reflection on one's solution after solving a problem in order to learn from it, giving up on a difficult problem after 10 minutes, preference for numerical vs. symbolic problems, and enjoyment in solving challenging physics problems.

Reliability refers to the relative degree of consistency between the survey scores, e.g., if an individual repeats the procedures [25-26]. One measure of reliability of a survey is the Cronbach's alpha ($\alpha_c$) which establishes the survey's reliability via internal consistency check. The Cronbach's alpha, $\alpha_c$, test was applied over all 33 questions for all groups (N=595), and $\alpha_c = 0.84$, which is reasonable from the standards of test-design. As noted later, there is very little variability in the responses for some of the groups (e.g., faculty) so it does not make sense to calculate $\alpha_c$ separately for the various groups [25-26].

Content validity refers to the degree to which the survey items reflect the domain of interest (in our case, attitudes and approaches to problem solving) [25-26]. As noted earlier, we discussed with some faculty members their opinions about productive approaches to problem solving and took their opinions into account while developing the additional survey questions. We further addressed the issue of content validity by taking steps to ensure that the respondents interpreted the survey questions as was intended. To this end we interviewed sample respondents from the introductory course, physics Ph.D. students (mostly those enrolled in a teaching assistant training course) and faculty members. During the interviews and discussions, we paid attention to respondents' interpretations of questions and modified the questions accordingly in order to make clear the actual intent of the questions. While the interviews with the introductory students were formal and audio-recorded, the discussions with the faculty members and Ph.D. students were informal and were not audio-recorded. The in-depth interviews with five introductory students from a first-semester algebra-based class, and discussions with the Ph.D. students and three physics faculty members helped modify the survey.

The interviews with the students from introductory class were particularly helpful in ensuring that the questions were interpreted clearly by the introductory students. Of approximately 40 introductory students responding to the invitation for paid one-on-one interviews with the researchers, five were selected. Since we wanted all students to be able to interpret the problems, two students were randomly chosen for interviews from those who scored above 70% and three students were chosen who obtained below 70% on their first midterm exam.

The survey questions were administered to all interviewed students and faculty members in the form of statements that they could agree or disagree with on a scale of 1 (strongly agree) to 5 (strongly disagree) with 3 signifying a neutral response. During the individual interviews with introductory students, students were also asked to solve some physics problems using a think-aloud protocol to gauge whether their responses to the survey questions about their attitudes and approaches to problem solving were consistent with the attitudes and approaches displayed while actually solving problems. In this protocol, we asked individuals to talk aloud while answering the questions. We did not disturb them while they were talking and only asked for clarifications of the points they did not make clear on their own later. While it is impossible to grasp all facets of problem solving fully by having students solve a few problems, a qualitative comparison of their answers to the survey questions and their actual approaches to solving problems was done after the interviews using the think aloud protocol. This comparison shows that students were consistent in their survey responses in many cases but in some instances they selected more favorable (expert-like) responses to the survey questions than the expertise that was explicitly evident from their actual problem solving. In this sense, the self-reported favorable responses (at least for the introductory students) should be taken as the upper limit of the actual favorable attitudes and approaches to problem solving.

We also tested validity of the survey by comparing actual survey data with those predicted according to the assumption of expert-novice behaviors, pre-defining the majority faculty response for each question as the "expert" response. We did not differentiate between "agree" and "strongly agree" in interpreting the data. Similarly, "disagree" and "strongly disagree" were combined for streamlining the data and their interpretation. A favorable response refers to either "agree" or "disagree" based upon which one was chosen by a majority of physics faculty. As we will discuss later, for most questions, the favorable response is supported by almost the entire faculty, but for a few questions the favorable response may only have the support of 60 – 70% of the faculty. In Table 1, we display data for individual questions for each statistical group as well as the average response for all 33 questions for each of the groups in the ``Avg." column. In the data reported in Table 1, the average score is as defined by Marx and Cummings [17]. To calculate the average score for a question, a +1 is assigned to each favorable response, a -1 is assigned to each unfavorable response, and a 0 is assigned to neutral responses. We then average these values for everybody in a particular group (e.g., introductory students) to obtain an average score for that group. Thus, the average score on each question for a group (which can vary between +1 and -1) indicates how expert-like the survey response of the group is on each survey question.

Table 1 shows that the faculty had close to unanimous agreement on most of the survey questions. These results support content validity of the survey. Table 1 also shows that faculty members, in general, answered the questions in a more expert-like fashion than Ph.D. students, who in turn were more expert-like than introductory-level students. While these differences cannot be quantified a-priori, such differences can be expected based upon the known expertise of each of these groups in physics. All these differences are statistically significant ($p<0.05$). These consistencies further provide validity to the survey.

To determine whether the differences between the groups are statistically significant and there is an appreciable effect size, we examined the groups as follows: all introductory students (we combined these classes since we did not find statistical differences between different introductory physics classes); all Ph.D. students and all faculty members. The effect sizes between groups over all 33 questions were calculated in the form of Cohen's d ($d= (\mu_1-\mu_2)/\sigma_{pooled}$), calculating individual group means (on a scale of -1 to +1) and standard deviations. Table 2 shows that the effect sizes between groups of different levels of expertise have a large to very large effect size ($1.19 < d < 2.18$), in favor of the more assumed expert-like group. Individual p-values for pairwise t-tests between each group shows that all differences between means are statistically significant. In other words, professors perform better than Ph.D. students who perform better than introductory students. Again these effect sizes are qualitatively consistent with the expected trends based upon the expertise of each group, and provide validity to the survey.

**Administration of the AAPS Survey**

After the validation of the survey, the final version of the AAPS survey was administered to several hundred introductory students in two different first-semester and second-semester algebra-based physics courses and to students in the first and second-semester calculus-based courses at a typical large state university in the US. Specifically, there were two sections of the first-semester algebra-based physics course with 209 students, two sections of the second-semester algebra-based physics course with 188 students, one first-semester calculus-based course section with 100 students, and a second-semester calculus-based course section with 44 students. In all of these courses, students were given a small number of bonus points for completing the survey.

The survey was also administered to 12 physics faculty who had taught introductory physics recently. Half of the faculty members were those who also gave the survey to their students. We also discussed faculty responses to selected questions individually with some of them. The expert (favorable) responses are given in the Appendix along with the survey. We also administered the final version of the survey to 24 Ph.D. candidates with the questions explicitly asking them to answer each question about their attitudes and approaches to introductory physics problem solving. We had individual discussions with four Ph.D. students about the reasoning for their AAPS survey responses and invited all Ph.D. students who had answered the survey questions to write a few sentences explaining their reasoning for selected survey questions online. Ten Ph.D. students who took the survey online provided written reasoning for their responses. An additional 18 Ph.D. candidates were administered the survey in the following year.

We report the data in two ways. First, the "net" average favorable response, shown in Table 1, was calculated as defined by Cummings et al. [18] as discussed earlier. A second method for representing the data is by separately showing the average percentage of favorable and unfavorable responses for each question for each group (the neutral responses are 100% minus the percentage of favorable and unfavorable responses). We will use this second method of data representation for all of our graphical representations of data.

**Results**

All of the data below for all groups pertains to attitudes and approaches to problem solving while solving introductory physics problems. Fig. 1 compares the survey responses of introductory students to 12 questions with the largest unfavorable (not expert-like) responses. The order of the questions is such that the unfavorable response is largest for the first question (question 20), second largest for the second question (question 12), etc. As shown in Fig. 1 (and also in Table 1), the most unfavorable response from the introductory students is on Question 20 implying many are not likely to take the time to reflect and learn from the problem solving after solving a homework problem. Their unfavorable responses in Fig. 1 to Questions 3, 5, 11 and 12 suggest that many introductory students believe that problem solving in physics basically means matching problems with the correct equations and then substituting values to get a number; being able to handle the mathematics is the most important part of the problem solving process; physics involves many equations, each of which applies primarily to specific situation; equations are not things that one needs to understand in an intuitive way; and they routinely use equations to calculate numerical answers even if they are non-intuitive. Also, their unfavorable responses to Questions 30 and 31 in Fig. 1 suggests that many students find it much more difficult to solve a physics problem with symbols than solving an identical problem with a numerical answer; and in problems with numerical answers they prefer to plug in numbers early on instead of solving those problems symbolically first. Moreover, their unfavorable responses to Questions 1 and 27 suggest that many introductory students feel stuck if they are not sure about how to start a problem, unless they get external help, and do not enjoy solving challenging physics problems. One introductory student in an interview noted that he feels frustrated with an incorrect problem solution and feels satisfied when he gets a problem right, since it motivates him to continue to do problem solving. Therefore, he likes easier problems. Also, in Fig. 1, introductory students' unfavorable responses to questions 9 and 16 suggest that many students do not use a similar approach to solve all problems involving the same physics principle, if the physical situations given in the problems are very different, and they mostly use their gut feeling rather than using physics principle to answer conceptual physics questions.

*Introductory Students: Comparison with Other Surveyed Groups*

Next, we compare introductory students' responses on selected questions on the AAPS survey with those of the physics faculty and Ph.D. students.

*Introductory students are still developing expertise*

Figure 2 shows that on Question 5 of the survey, while no faculty agreed with the statement (no unfavorable response) that problem solving in introductory physics basically means matching problems with the correct equations and then substituting values to get a number, the average responses of the introductory students and Ph.D. students are indistinguishable. But individual discussions show that there is difference in the reasoning of many introductory students and Ph.D. students. For example, many Ph.D. students felt so comfortable with the applications of basic principles that not much explicit thought was involved in solving the introductory level problems. On the other hand, many introductory students think that physics is a collection of disconnected facts and formulas and use a "plug and chug" approach without thinking if a principle is applicable in a particular context [13].

Figure 3 shows that, in response to Question 6, all of the physics faculty noted that while solving introductory physics problems they could often tell when their work and/or answer is wrong even without external resources but only approximately 50% and 80% of the introductory students and Ph.D. students, respectively, could do so.

Figure 4 shows that, in response to Question 11 about whether equations must be intuitive in order to use them or whether they routinely use equations even if they are non-intuitive, approximately 75% of faculty and Ph.D. students disagreed with the statements (favorable response). In contrast, only approximately 30% of the introductory students provided favorable response and the responses of the majority of introductory students show that they are likely to use equations to calculate answers even if they are non-intuitive (see Figure 4). This finding is consistent with the prior results that show that many introductory students view problem solving in physics as an exercise in finding the relevant equations rather than focusing on why a particular physics principle may be involved and building an intuition about a certain type of physics problems [13].

Figure 5 shows that, in response to Question 12 regarding whether physics involves many equations each of which applies primarily to a specific situation, all but one physics faculty disagreed with the statement (favorable). However, the percentage of introductory students and Ph.D. students who disagreed with the statement was slightly more than 35% and 80%, respectively. These responses are commensurate with the expertise of each group and the fact that experts are more likely to discern the coherence of the knowledge in physics and appreciate how very few laws of physics are applicable in diverse situations and can explain different physical phenomena.

In response to Question 25 about whether individuals make sure they learn from their mistakes and do not make the same mistakes again, all but one physics faculty members agreed with the statement (favorable) and one was neutral. On the other hand, only slightly more than 70% of the introductory students and Ph.D. students agreed with the statement. One introductory student said he did not review errors on the midterm exam as much as he would on homework, partly because the homework problems may show up on a future test but partly because he didn't like staring at his bad exam grade. The reluctance to reflect upon tests is consistent with our earlier findings, which demonstrated that many students did not reflect automatically on their mistakes for similar reasons [27-31].

Manipulation of symbols rather that numbers increases the difficulty of a problem for many introductory students. Question 30 asked whether symbolic problems were more difficult than identical problems with numerical answers and question 31 asked if individuals preferred to solve a problem with a numerical answer symbolically first and only plug in the numbers at the very end. Figures 6 and 7 show that the responses of physics faculty and Ph.D. students are comparable to each other but introductory students' responses are very different. Only approximately 35% of the introductory students disagreed with the statement that it is more difficult to solve a problem symbolically and 45% agreed with the statement that they prefer to solve the problem symbolically first and only plug in the numbers at the very end.

Individual discussions with some introductory students show that they have difficulty keeping track of the variables they are solving for if they have several symbols to consider, which motivates them to substitute numbers at the beginning of the solutions. One strategy to help introductory students feel more confident about using symbols is to ask them to underline the variable they are solving for so as to keep it from getting mixed with the other variables. Some introductory students noted that they did not like carrying expressions involving symbols from one equation to another because they were afraid that they would make mistakes in simplifying the expressions. Developing mathematical facility can help students develop the confidence to solve the problems symbolically first before substituting values. In addition, instructors should help students understand why it is useful to keep the symbols till the end, including the fact that it can allow students to check the solution, e.g., by checking the dimension, and it can also allow them to check the limiting cases useful for developing confidence in one's solution.

Figure 8 shows that, in response to Question 2 about whether they often make approximations about the physical world when solving introductory physics problems, all faculty members noted that they do so. However, less than half of the introductory students and about two thirds of the Ph.D. students noted they do so. Individual discussions with some faculty showed that they considered the idealization of the problem in introductory physics (e.g., framing problems without friction or air resistance, considering spherical cows or point masses, etc.) as making approximations about the physical world and they felt that such approximations were helpful for getting an analytical answer and for building intuition about physical phenomena. It appears that students who noted that they don't make approximations may not have carefully thought about the role of approximations about the physical world in physics problem solving.

*Responses to some questions may have unexpected trends*

For some survey questions, faculty responses were not unanimous and may even look similar to the responses of introductory students on average. These responses should be interpreted by carefully identifying the reasoning for the responses for each group. For example, Figure 9 shows that, in response to Question 14 regarding whether they always explicitly think about concepts that underlie the problems when solving introductory physics problems, close to 90% of the Ph.D. students agreed (favorable) that they do so but only approximately 65% and 55% of the physics faculty and introductory students agreed, respectively. The non-monotonic nature of the responses in Figure 9 going from the introductory students to faculty may seem surprising at first, but individual discussions show that some faculty do not always explicitly think about the concepts that underlie the problem because the concepts have become obvious to them due to their vast experience. They are able to invoke the relevant physics principles, e.g., conservation of mechanical energy or conservation of momentum, automatically when solving an introductory problem without making a conscious effort. In contrast, introductory students often do not explicitly think about the relevant concepts because they often consider physics as consisting of disconnected facts and formulas and associate physics problem solving as a task requiring looking for the relevant formulas without performing a conceptual analysis and planning of the problem solution [13]. Thus, the reasoning behind the less favorable responses of faculty to Question 14 is generally very different from the reasons behind the introductory students' responses.

Problem solving is often a missed learning opportunity because, in order to learn from problem solving, one must reflect upon the problem solution [27-31]. For example, one must ask questions such as "what did I learn from solving this problem?", "why did the use of one principle work and not the other one?" or "how will I know that the same principle should be applicable when I see another problem with a different physical situation?" Unfortunately, the survey results show a general lack of reflection by individuals in each group after solving problems.

Figure 10 shows that in response to Question 20, only approximately 25% of introductory students noted that they reflect and learn from problem solutions. Since reflection is so important for learning and building a robust knowledge structure, these findings suggest that instructors should consider giving students explicit incentive to reflect after they solve physics problems [27-31]. Moreover, only approximately 55% of the Ph.D. students and about 75% of faculty noted that, they take the time to reflect and learn from the solution to introductory physics problems. Individual discussions show that, for introductory level problems, both physics faculty and Ph.D. students felt that they monitor their thought processes while solving the problems since the problems are relatively simple. Therefore, reflection at the end of problem solving is not required.

Moreover, in response to Question (24), 63% of introductory students, 54% of Ph.D. students and 67% of the faculty noted that they like to think through a difficult physics problem with a peer. Individual discussions with some of the faculty members and students suggested that the choice of whether one continues to persevere individually or works with a peer to solve a challenging problem depends on an individual's personality. While many from each group agree that talking to peers may be helpful, at least for challenging problems, some of them are inherently more averse to discussions with peers than others.

Figure 11 shows that, in response to Question 3, regarding whether being able to handle the mathematics is the most important part of the process in solving an introductory physics problem, less than 60% of the faculty were favorable and disagreed with the statement and approximately 35% were neutral. Also, more introductory students provided favorable responses compared to the Ph.D. students. Individual discussions with some faculty about Question 3 suggest that they felt that conceptual knowledge in physics was the central aspect of physics problem solving. But those who were neutral in response to Question 3 felt that the students would not excel in physics without a good grasp of mathematics even though concepts are important. Thus, Question 3 is one question for which there isn't a strong faculty agreement because some felt that both mathematics and physical concepts were vital to problem solving.

Similarly, in response to Question 16, 50% of the introductory students claimed that they use their "gut" feeling to answer conceptual questions rather than invoking physics principles. Introductory students often view conceptual questions as guessing tasks and use their "gut" feeling rather than explicitly considering how the physical principles apply in those situations [13, 32-33]. One interviewed introductory student stated that he would not consider principles when answering a conceptual question because over-analyzing the problem is more likely to make his answer wrong. When Mazur from Harvard University gave the Force Concept Inventory Conceptual standardized test [34] to his introductory students, a student asked if he should do it the way he really thinks about it or the way he has been taught to think about it [33]. It appears that students sometimes hold two views simultaneously, one based upon their gut feeling and another is based upon what they learned in the physics class, and their views coexist and are difficult to merge. Moreover, only 75% of faculty (and an even smaller percentage for Ph.D. students) noted that, while answering conceptual physics questions, they use the physics principles they usually think about when solving quantitative problems rather than mostly using their "gut" feeling. Discussions elucidated that the faculty members' use of their "gut" feeling to answer conceptual questions (rather than explicitly invoking physics principles) was often due to the fact that they had developed good intuition about the problems based upon their vast experience [35]. They did not need to explicitly think about the physical principles involved.

*Factor Analysis*

Adams and Wieman [23] list the pros and cons of confirmatory factor analysis (CFA) and exploratory factor analysis (EFA). While CFA is appealing from the point of view that it starts with categorizations based upon expert views about how the items should be grouped, a major difficulty with CFA is that the expert categorization of the items does not always fit with the way students actually categorize items. We performed a principal component analysis (PCA), which uses empirical data to find the relationships and patterns among AAPS survey items [25-26]. The data from 541 introductory students, which easily meets the recommended sample size of at least 300 participants, was used to group the items using PCA. To check the suitability of our data for PCA, we used the Kaiser-Meyer-Olkin (KMO) measure of sampling adequacy, and Bartlett's sphericity test to confirm that our example has patterned relationships. The KMO value was found to be 0.845, which established that the extracted factors in the PCA account for most of the variance in responses, and Bartlett's test was found to be statistically significant ($\chi^2$=1733.34, df=528, p=.000), which established that the original variables were strongly correlated. Since all requirements are met, distinct and reliable factors can be expected from our sample.

To maximize independence of variables and percentage of variance explained in the PCA, a Promax oblique axis rotation was performed. A total of 9 factors with eigenvalues of at least 1 (composing of groups of at least two survey items) were obtained that explained 53% of the total variance. The variance explained by the scale indicates that the AAPS survey measured the students' attitudes and approaches to physics problem solving adequately. All questions on the survey appear at least once in these nine factors. Thus, all AAPS survey items are likely to make a meaningful contribution to the survey.

Table 3 presents the findings of the exploratory factor analysis, in which each factor has a description that summarizes the common link between the questions within that factor. The researchers came up with the descriptions separately and then they discussed the descriptions and jointly agreed on the descriptions after discussions. Some of the factors focus on attitudes and approaches to problem solving in specific cases (e.g., drawing diagrams/pictures and doing scratch work) while others focus on boarder issues.

**Summary and Conclusions**

We developed, validated and administered the AAPS survey to introductory students and compared their responses to those of physics faculty members and Ph.D. students. We discussed the responses individually with some students and faculty to improve the item wording and understand the rationale for the responses.

The responses of introductory students on the survey were often less favorable than faculty and Ph.D. students. For example, unlike the introductory students, all physics faculty members noted that they enjoy solving challenging physics problems. We also find that on some survey questions, the introductory students' and faculty members' responses to the survey questions must be interpreted carefully. For example, only two thirds of the faculty noted that they always think about the concepts that underlie the problem explicitly while solving problems. Individual discussions with faculty members suggests that they felt that, after years of teaching experience, the concepts that underlie many of the problems have become "automatic" for them and they do not need to explicitly think about them. The fact that many introductory students always think explicitly about the concepts that underlie the problems while solving problems suggests that they have not developed the same level of expertise and efficiency in solving problems as physics faculty have.

Comparison of introductory students' survey responses with Ph.D. students' responses shows that, in general, Ph.D. students have more favorable attitudes and approaches to solving introductory physics problems due to their higher level of expertise than the introductory students. However, on some questions, the responses must be interpreted carefully in light of the explanations provided by the students. For example, in response to whether the problem solving in physics is essentially "plug and chug," the average response of the introductory students and Ph.D. students is indistinguishable. Interviews and written responses suggest that Ph.D. students have developed sufficient expertise in introductory physics so that solving such problems does not require much explicit thought and they can often immediately tell which principle of physics is applicable in a particular situation. On the other hand, many introductory students forgo the conceptual analysis and planning of the problem solution and immediately look for the formulas when they should not [2,36-38].

Also, survey responses and individual discussions suggest that compared to the introductory students, Ph.D. students find introductory physics equations more intuitive and are better able to discern the applicability of a physics principle epitomized in the form of a mathematical equation to diverse situations due to their higher level of expertise. In solving introductory physics problems, the fraction of Ph.D. students who noted that they reflect and learn from the problem solution after solving a problem is significantly larger than the fraction of introductory students who noted doing so in their courses. Also, many students did not reflect on the exam solutions even after they received the solutions because they did not expect those problems to show up again on another exam. There was a large difference between the introductory students' and Ph.D. students' responses in their facility to manipulate symbols (vs. numbers) with introductory students finding it more difficult to solve problems given in symbolic form. In problems where numbers were provided, many introductory students noted that they prefer to plug numbers early on rather than waiting till the end to do so as an expert would do.

In general, the less favorable responses of the introductory students on the survey compared to those of the faculty and Ph.D. students imply that introductory students have less expertise than physics faculty and Ph.D. students. While one can rationalize these unfavorable responses of introductory students by noting that they do not have as much experience solving physics problems, instruction should explicitly focus on helping them learn effective approaches and attitudes to problem solving while developing problem solving and meta-cognitive skills and learning physics concepts. This may also help students develop a broader conception of what it means to learn physics [3-8]. Moreover, instructors should include both quantitative and conceptual questions to motivate students to reflect on the problem solving process and to help them develop intuition about the equations underlying the problems.

**Acknowledgements**
We thank all of the faculty and students at various levels who helped during the validation of the survey. We also thank Fred Reif and Chris Schunn for extremely helpful discussions. C.S. was supported by the US National Science Foundation award NSF-PHY-1505460 and NSF-PHY-1202909.

**Appendix: AAPS Survey and Favorable (Expert-like) Responses**

To what extent do you agree with each of the following statements when you solve physics problems?

Answer with a single letter as follows:
A) Strongly Agree
B) Agree Somewhat
C) Neutral or Don't Know
D) Disagree Somewhat
E) Strongly Disagree

1. If I'm not sure about the right way to start a problem, I'm stuck unless I go see the teacher/TA or someone else for help.

2. When solving physics problems, I often make approximations about the physical world.

3. In solving problems in physics, being able to handle the mathematics is the most important part of the process.

4. In solving problems in physics, I always identify the physics principles involved in the problem first before looking for corresponding equations.

5. "Problem solving" in physics basically means matching problems with the correct equations and then substituting values to get a number.

6. In solving problems in physics, I can often tell when my work and/or answer is wrong, even without looking at the answer in the back of the book or talking to someone else about it.

7. To be able to use an equation to solve a problem (particularly in a problem that I haven't seen before), I think about what each term in the equation represents and how it matches the problem situation.

8. There is usually only one correct way to solve a given problem in physics.

9. I use a similar approach to solving all problems involving conservation of linear momentum even if the physical situations given in the problems are very different.

10. If I am not sure about the correct approach to solving a problem, I will reflect upon physics principles that may apply and see if they yield a reasonable solution.

11. Equations are not things that one needs to understand in an intuitive sense; I routinely use equations to calculate numerical answers even if they are non-intuitive.

12. Physics involves many equations each of which applies primarily to a specific situation.

13. If I used two different approaches to solve a physics problem and they gave different answers, I would spend considerable time thinking about which approach is more reasonable.

14. When I solve physics problems, I always explicitly think about the concepts that underlie the problem.

15. When solving physics problems, I often find it useful to first draw a picture or a diagram of the situations described in the problems.

16. When answering conceptual physics questions, I mostly use my "gut" feeling rather than using the physics principles I usually think about when solving quantitative problems.

17. I am equally likely to draw pictures and/or diagrams when answering a multiple-choice question or a corresponding free-response (essay) question.

18. I usually draw pictures and/or diagrams even if there is no partial credit for drawing them.

19. I am equally likely to do scratch work when answering a multiple-choice question or a corresponding free-response (essay) question.

20. After I solve each physics homework problem, I take the time to reflect and learn from the problem solution.

21. After I have solved several physics problems in which the same principle is applied in different contexts, I should be able to apply the same principle in other situations.

22. If I obtain an answer to a physics problem that does not seem reasonable, I spend considerable time thinking about what may be wrong with the problem solution.

23. If I cannot solve a physics problem in 10 minutes, I give up on that problem.

24. When I have difficulty solving a physics homework problem, I like to think through the problem with a peer.

25. When I do not get a question correct on a test or homework, I always make sure I learn from my mistakes and do not make the same mistakes again.

26. It is more useful for me to solve a few difficult problems using a systematic approach and learn from them rather than solving many similar easy problems one after another.

27. I enjoy solving physics problems even though it can be challenging at times.

28. I try different approaches if one approach does not work.

29. If I realize that my answer to a physics problem is not reasonable, I trace back my solution to see where I went wrong.

30. It is much more difficult to solve a physics problem with symbols than solving an identical problem with a numerical answer.

31. While solving a physics problem with a numerical answer, I prefer to solve the problem symbolically first and only plug in the numbers at the very end.

32. Suppose you are given two problems. One problem is about a block sliding down an inclined plane with no friction present. The other problem is about a person swinging on a rope. Air resistance is negligible. You are told that both problems can be solved using the concept of conservation of mechanical energy of the system. Which one of the following statements do you MOST agree with? (Choose only one answer.)
    A) The two problems can be solved using very similar methods.
    B) The two problems can be solved using somewhat similar methods.
    C) The two problems must be solved using somewhat different methods.
    D) The two problems must be solved using very different methods.
    E) There is not enough information given to know how the problems will be solved.

33. Suppose you are given two problems. One problem is about a block sliding down an inclined plane. There is friction between the block and the incline. The other problem is about a person swinging on a rope. There is air resistance between the person and air molecules. You are told that both problems can be solved using the concept of conservation of total (not just mechanical) energy. Which one of the following statements do you MOST agree with? (Choose only one answer.)
    A) The two problems can be solved using very similar methods.
    B) The two problems can be solved using somewhat similar methods.
    C) The two problems must be solved using somewhat different methods.
    D) The two problems must be solved using very different methods.
    E) There is not enough information given to know how the problems will be solved.

*Favorable Responses for AAPS by Question*

1. D/E
2. A/B
3. D/E
4. A/B
5. D/E
6. A/B
7. A/B
8. D/E
9. A/B
10. A/B
11. D/E
12. D/E
13. A/B
14. A/B
15. A/B
16. D/E
17. A/B
18. A/B
19. A/B
20. A/B
21. A/B
22. A/B
23. D/E
24. A/B
25. A/B
26. A/B
27. A/B
28. A/B
29. A/B
30. D/E
31. A/B
32. A/B
33. A/B

# Tables

**Table 1.** Average responses for introductory students, Ph.D. students and faculty for each question, and averaged over all survey questions (see the last entry).

| Problem number | 1 | 2 | 3 | 4 | 5 | 6 | 7 |
|---|---|---|---|---|---|---|---|
| Introductory students | 0.14 | 0.19 | 0.15 | 0.41 | 0.16 | 0.24 | 0.61 |
| Ph.D. students | 0.71 | 0.42 | -0.04 | 0.83 | 0.17 | 0.75 | 0.83 |
| Faculty | 0.83 | 1.00 | 0.50 | 0.92 | 0.92 | 1.00 | 0.83 |
| Problem number | 8 | 9 | 10 | 11 | 12 | 13 | 14 |
| Introductory students | 0.67 | 0.24 | 0.58 | -0.03 | -0.06 | 0.56 | 0.32 |
| Ph.D. students | 0.83 | 0.46 | 0.88 | 0.67 | 0.54 | 0.88 | 0.88 |
| Faculty | 0.92 | 0.58 | 0.92 | 0.67 | 0.83 | 1.00 | 0.50 |
| Problem number | 15 | 16 | 17 | 18 | 19 | 20 | 21 |
| Introductory students | 0.74 | 0.23 | 0.55 | 0.69 | 0.77 | -0.19 | 0.71 |
| Ph.D. students | 0.96 | 0.50 | 0.79 | 0.96 | 0.88 | 0.38 | 0.92 |
| Faculty | 1.00 | 0.67 | 1.00 | 0.92 | 1.00 | 0.75 | 1.00 |
| Problem number | 22 | 23 | 24 | 25 | 26 | 27 | 28 |
| Introductory students | 0.52 | 0.40 | 0.43 | 0.56 | 0.37 | 0.03 | 0.75 |
| Ph.D. students | 1.00 | 1.00 | 0.21 | 0.54 | 0.71 | 0.67 | 0.96 |
| Faculty | 1.00 | 0.92 | 0.42 | 0.92 | 1.00 | 0.92 | 1.00 |
| Problem number | 29 | 30 | 31 | 32 | 33 | All | |
| Introductory students | 0.74 | -0.04 | 0.08 | 0.70 | 0.46 | 0.38 | |
| Ph.D. students | 1.00 | 0.92 | 0.92 | 1.00 | 0.83 | 0.72 | |
| Faculty | 1.00 | 1.00 | 1.00 | 1.00 | 1.00 | 0.88 | |

**Table 2.** Effect sizes calculated using Cohen's d between introductory students, physics Ph.D. students, and faculty regarding introductory-level physics problem-solving on the AAPS survey. Sample sizes for different groups were accounted for in calculating Cohen's d.

| | Cohen's d | |
|---|---|---|
| Group | Ph.D. students | Faculty |
| Intro physics | 1.60 | 2.18 |
| Ph.D. students | | 1.19 |
| Faculty | | |

**Table 3.** Principal component analysis results featuring the 9 primary factors (eigenvalues greater than 1), and descriptions of commonalities between questions for each factor.

| Factor (% of variance explained) | Item | Loading | Description |
|---|---|---|---|
| Factor1 (13) | 13 | .77 | Metacognition and enjoyment in physics problem solving |
| | 14 | .73 | |
| | 7 | .70 | |
| | 10 | .69 | |
| | 22 | .63 | |
| | 29 | .60 | |
| | 20 | .51 | |
| | 4 | .57 | |
| | 25 | .38 | |
| | 27 | .27 | |
| | 6 | .28 | |
| | 21 | .29 | |
| Factor2 (8) | 18 | .90 | Utility of pictures, diagrams or scratch work in physics problem solving |
| | 17 | .88 | |
| | 15 | .75 | |
| | 19 | .65 | |
| Factor3 (6) | 5 | .69 | Perception of problem solving approach |
| | 11 | .68 | |
| | 12 | .55 | |
| | 8 | .33 | |
| | 26 | .28 | |
| | 9 | -.28 | |
| Factor4 (5) | 8 | .76 | General expert-novice differences in physics problem solving |
| | 28 | .56 | |
| | 21 | .43 | |
| | 24 | .43 | |
| | 29 | .30 | |
| Factor5 (5) | 31 | .84 | Difficulty in solving problems symbolically |
| | 30 | .83 | |
| Factor6 (5) | 1 | .75 | Problem solving confidence |
| | 24 | -.71 | |
| | 23 | .50 | |
| | 6 | .30 | |
| Factor7 (5) | 33 | .88 | Solving different problems using the same principle |
| | 32 | .80 | |
| Factor8 (4) | 16 | .76 | Sense-making |
| | 2 | -.56 | |
| | 5 | .30 | |
| Factor9 (3) | 3 | .77 | Problem solving sophistication |
| | 20 | -.35 | |
| | 25 | -.32 | |
| | 9 | .29 | |

# Figures

**Figure 1.** Comparison of introductory students' survey responses to 12 selected questions. The order of the questions is such that the unfavorable response is largest for the first question (Question 20), second largest for the second question (Question 12), etc. Uncertainty in data [25] was interpreted in terms of standard error (SE). Typical SE values for introductory students on each of the AAPS items were between +/-0.03 and +/-0.04, with the average standard error of +/-0.033.

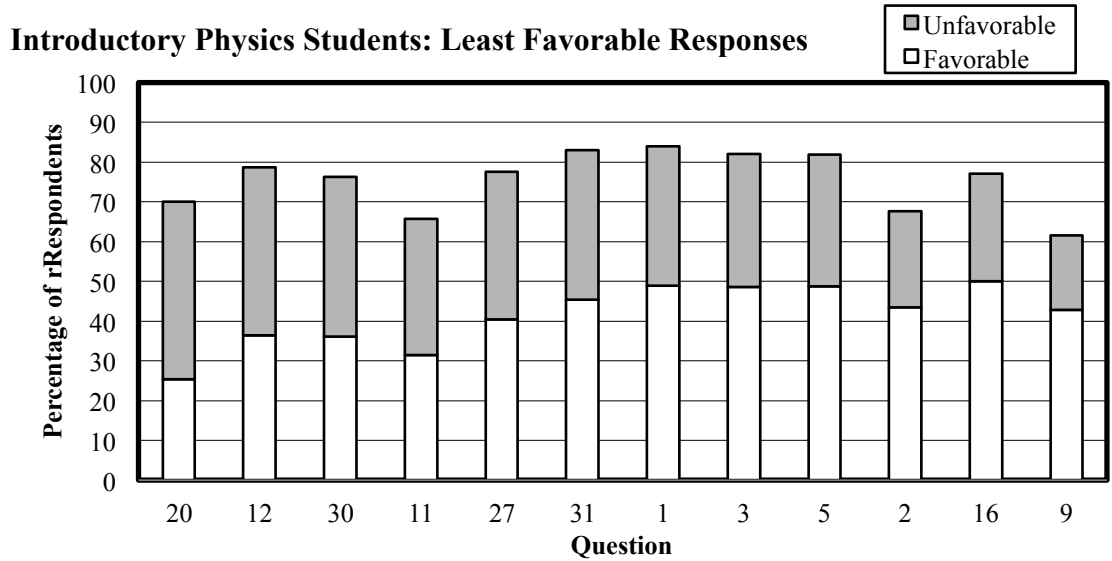

**Figure 2.** Histogram showing favorable (disagree) and unfavorable (agree) responses for survey question 5 about whether problem solving in physics is mainly an exercise in finding the right equation/formula. The histogram shows that a large number of non-faculty respondents from all groups agreed with the statement or were neutral. The SE was +/-0.04 for introductory students, +/-0.18 for Ph.D. students, and +/-0.08 for faculty.

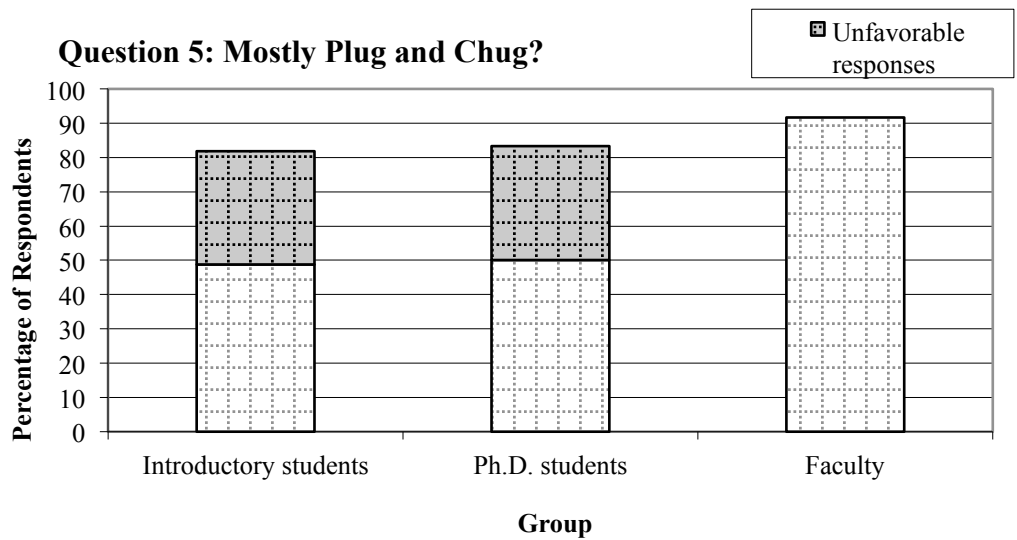

**Figure 3.** Histogram showing favorable (agree) and unfavorable (disagree) responses for survey Question 6. The histograms show that faculty members were always aware of when they were wrong in problem solving but other respondents were less certain. Only slightly more than 50% of introductory students could tell that their answers were wrong in physics problem solving. The SE was +/-0.04 for introductory students and +/-0.11 for Ph.D. students; faculty were unanimous and did not have any variance.

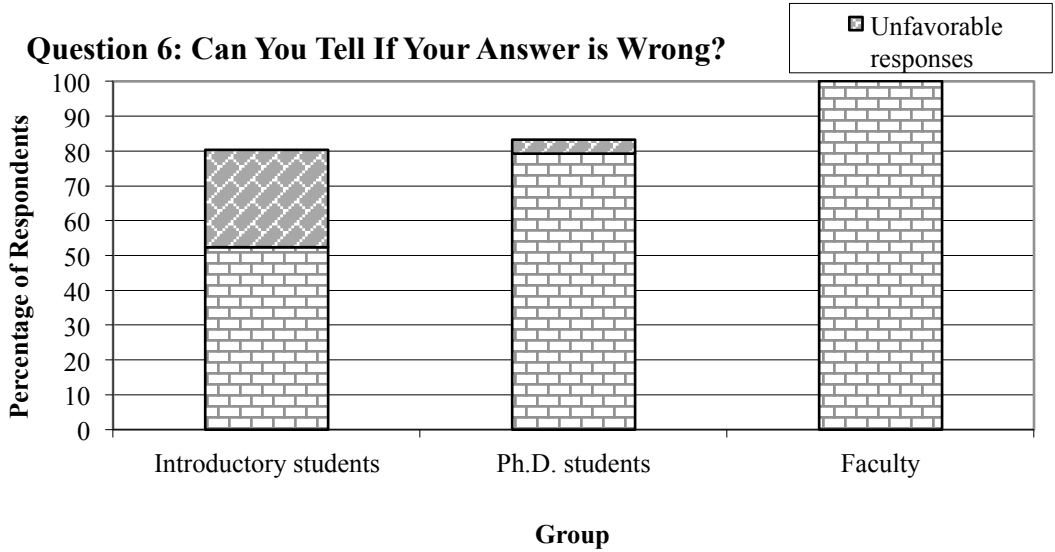

**Figure 4.** Histogram showing favorable (disagree) and unfavorable (agree) responses for survey Question 11. The histogram shows that an almost equal number of introductory students agreed or disagreed with the statement or where neutral. The SE was +/-0.03 for introductory students, +/-0.13 for Ph.D. students, and +/-0.18 for faculty.

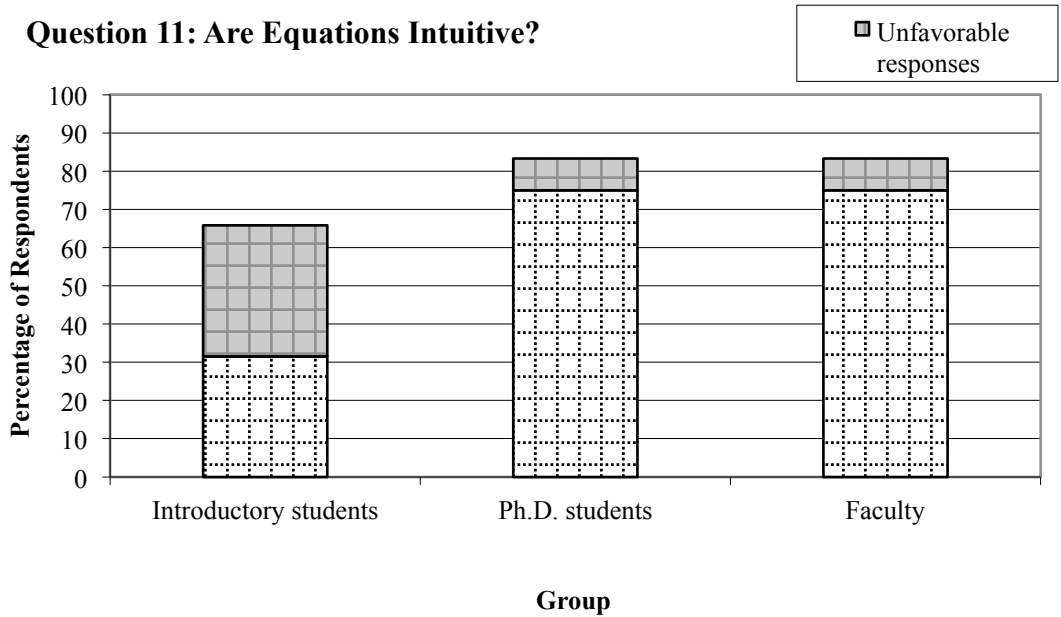

**Figure 5.** Histogram showing favorable (disagree) and unfavorable (agree) responses for survey Question 12 about whether physics involves many equations that each apply primarily to a specific situation. As we go from the introductory students to the faculty, the disagreement with the statement increases. The SE was +/-0.04 for introductory students, +/-0.17 for Ph.D. students, and +/-0.16 for faculty.

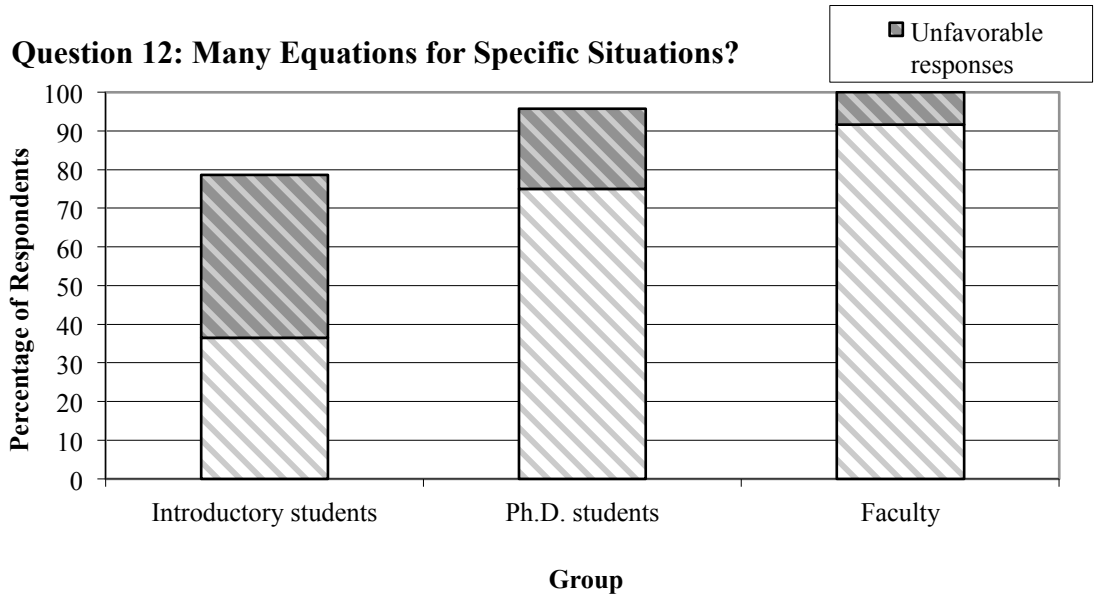

**Figure 6.** Histogram showing favorable (disagree) and unfavorable (agree) responses for survey Question 30. The histogram shows that faculty and Ph.D. students did not believe that it is more difficult to solve a physics problem with symbols than solving an identical problem with numerical answer but introductory students often did. The SE was +/-0.04 for introductory students and +/-0.08 for Ph.D. students. Faculty were unanimous and had no variance.

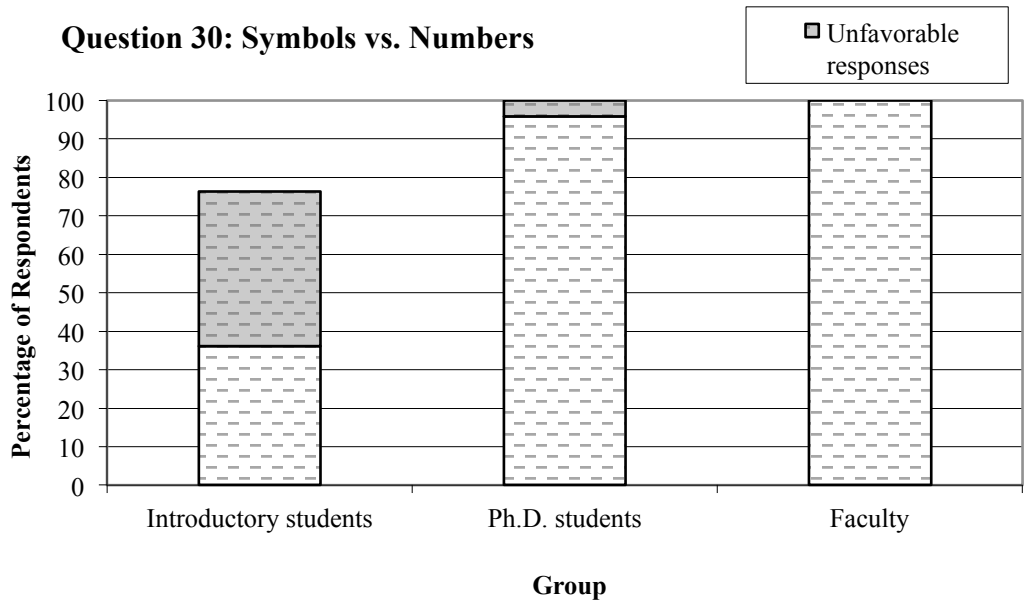

**Figure 7.** Histogram showing favorable (agree) and unfavorable (disagree) responses for survey Question 31. The histogram shows that faculty and Ph.D. students preferred to solve a problem symbolically first and only plug in the numbers at the very end but less than half of the introductory students agreed with them. The SE was +/-0.04 for introductory students and +/-0.08 for Ph.D. students. Faculty were unanimous and had no variance.

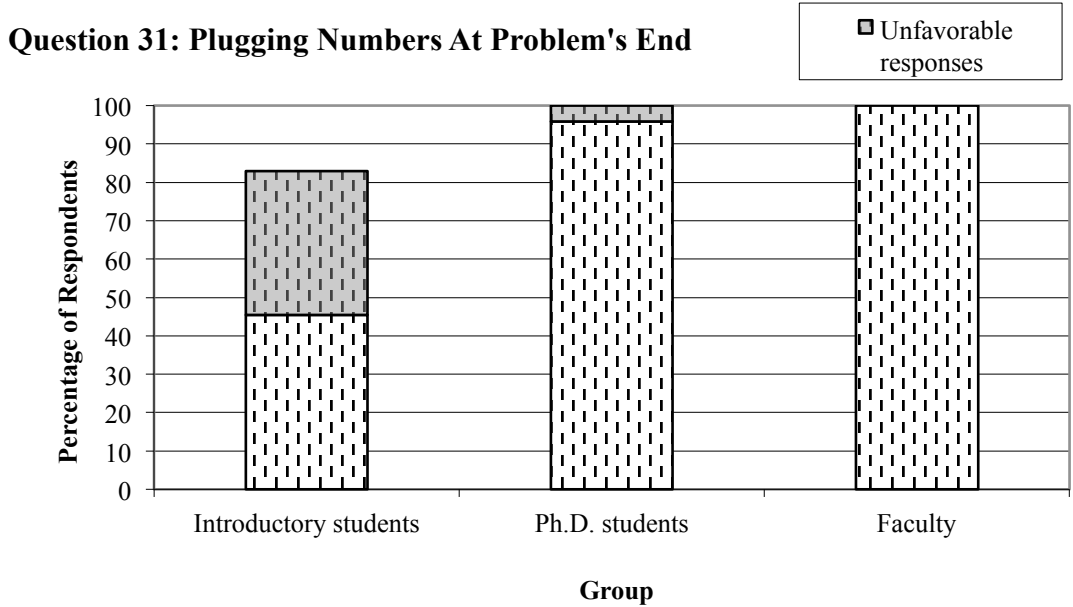

**Figure 8.** Histogram showing favorable (agree) and unfavorable (disagree) responses for survey Question 2. The histogram shows that all faculty members agreed that they often make approximations about the physical world but less than half of the introductory students and two thirds of the physics Ph.D. students were in agreement. The SE was +/-0.03 for introductory students and +/-0.18 for Ph.D. students. Faculty were unanimous and had no variance.

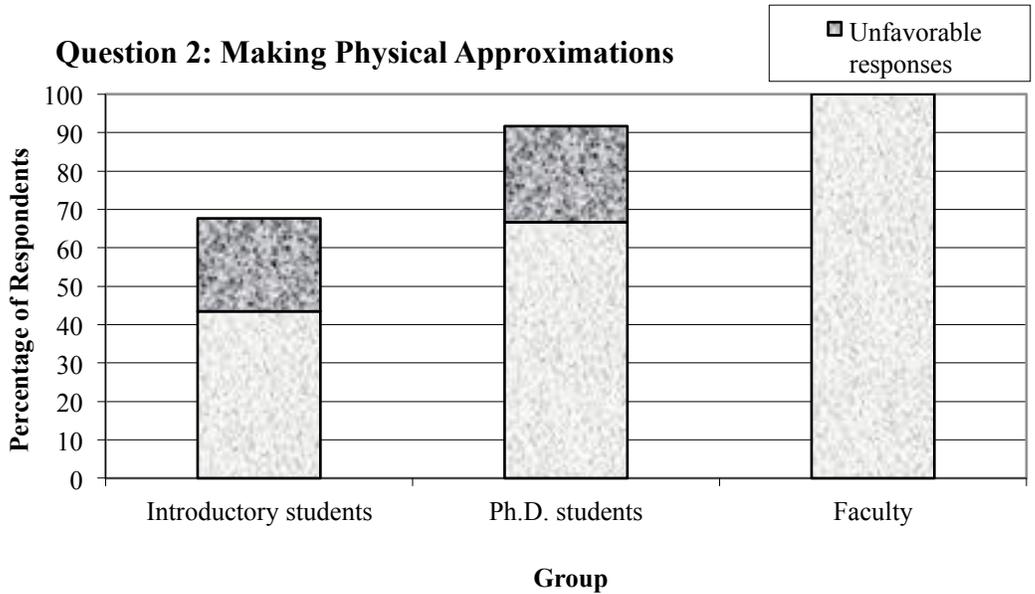

**Figure 9.** Histogram showing favorable (agree) and unfavorable (disagree) responses for survey Question 14. While the non-monotonic trend in favorable responses from introductory students to faculty may seem surprising, some faculty noted that they do not explicitly think about concepts that underlie the problem while solving problems because the concepts have become obvious to them whereas introductory students often do not think about concepts because they believe in a plug and chug approach. The SE was +/-0.04 for introductory students, +/-0.07 for Ph.D. students, and +/-0.22 for faculty members.

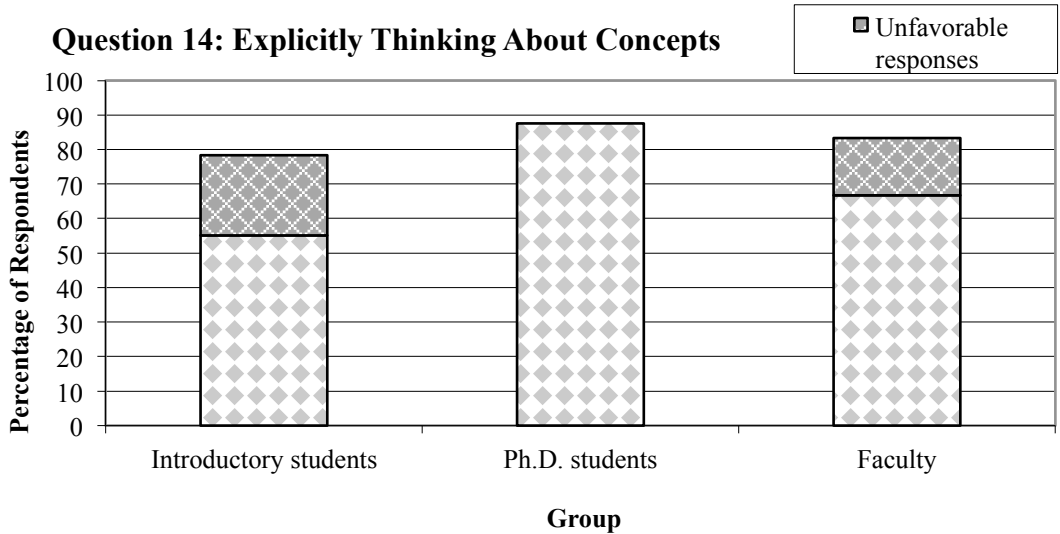

**Figure 10.** Histogram showing favorable (agree) and unfavorable (disagree) responses for survey Question 20. The histogram shows that none of the groups had 80% individuals who agreed that they take the time to reflect and learn from the problem solutions after solving problems but the reasons for the lack of reflection varied across different groups. The SE was +/-0.035 for introductory students, +/-0.15 for Ph.D. students, and +/-0.125 for faculty members.

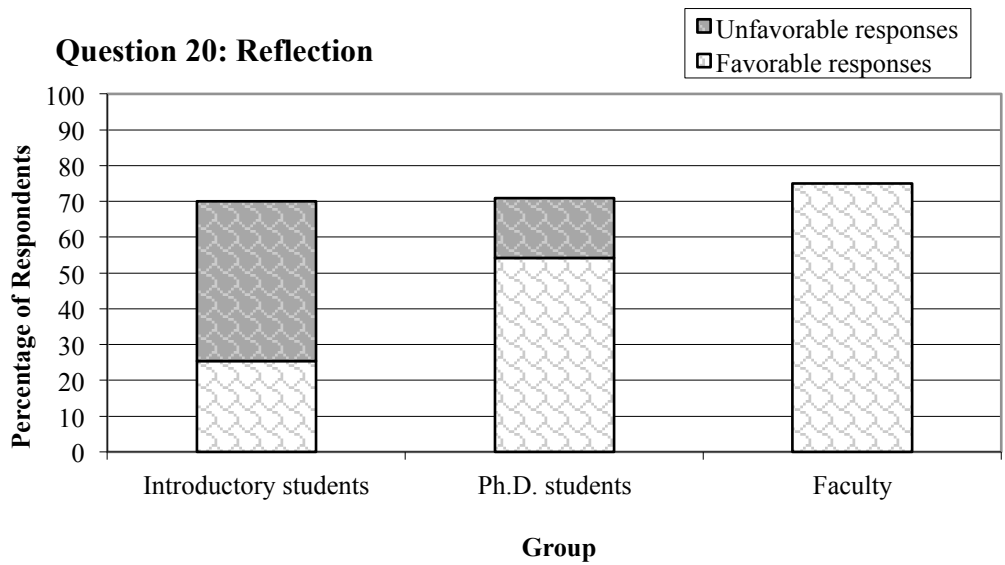

**Figure 11.** Histogram showing favorable (disagree) and unfavorable (agree) responses for survey Question 3 about whether mathematics is the most important part of the problem solving process. The histograms show that a large number of introductory students and Ph.D. students agreed with the statement. The SE was +/-0.04 for introductory students, +/-0.18 for Ph.D. students, and +/-0.19 for faculty members.

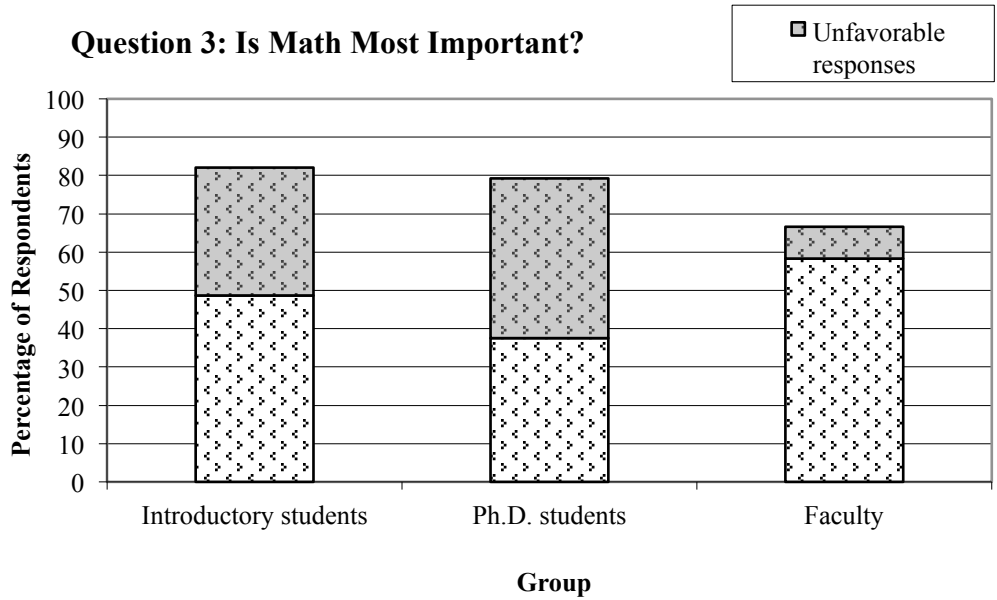